\documentclass[aps,twocolumn,showpacs, 10pt]{revtex4-1}
\usepackage{graphicx,color}
\usepackage{amsmath}
\usepackage{amssymb}
\usepackage{amsfonts}
\usepackage{bm}
\usepackage{hyperref}
\usepackage{float}
\usepackage{color}

\newcommand{\ot}{{\,\otimes\,}}
\newcommand{{\Cd}}{{\mathbb{C}^d}}

\def\oper{{\mathchoice{\rm 1\mskip-4mu l}{\rm 1\mskip-4mu l}%
{\rm 1\mskip-4.5mu l}{\rm 1\mskip-5mu l}}}
\def\<{\langle}
\def\>{\rangle}
\newtheorem{Theorem}{Theorem}

\newtheorem{Remark}{Remark}
\newtheorem{Example}{Example}

\begin{document}

\title{Entanglement witnesses from mutually unbiased bases}
\author{Dariusz Chru\'sci\'nski$^1$, Gniewomir Sarbicki$^1$ and Filip Wudarski$^{1,2,3,4}$\\
$^1$Institute of Physics, Faculty of Physics, Astronomy and Informatics, \\ Nicolaus Copernicus University,
Grudzi\c{a}dzka 5/7, 87--100 Toru\'n, Poland \\
$^2$ Quantum Research Group, School of Chemistry and Physics,
University of KwaZulu-Natal, Durban 4001, South Africa,\\
$^3$ National Institute for Theoretical Physics (NITheP), KwaZulu-Natal, South Africa \\
$^4$Physikalisches Institut, Universit\"at Freiburg, Herrmann-Herder-Stra{\ss}e 3, D-79104 Freiburg, Germany}

\begin{abstract}
We provide a class of entanglement witnesses constructed in terms of Mutually Unbiased Bases (MUBs). This construction reproduces many well-known examples like the celebrated reduction map and Choi map together with its generalizations. We illustrated our construction by the detailed analysis of the 3-dimensional case: in this case, one  obtains a family of entanglement witnesses parameterized by an $L$-dimensional torus ($L=2,3,4$ being a number of MUBs used in the construction).
\end{abstract}


\maketitle

\section{Introduction}

Quantum entanglement is one of the most fundamental resources for modern quantum technologies and quantum information processing \cite{QIT,HHHH,Guhne}. It is therefore clear the characterization of entanglement and other quantum correlations \cite{Modi} is of great importance for quantum information science (actually, the problem of determining  whether or not a given state is entangled is NP-hard \cite{NP}). For low dimensional systems $\mathbb{C}^2 \ot \mathbb{C}^2$ (qubit-qubit) and $\mathbb{C}^2 \ot \mathbb{C}^3$ (qubit-qutrit) the problem is solved due to the celebrated Peres-Horodecki partial transposition criterion \cite{4,5}. However, for more complex systems there are states passing the partial transposition criterion (so called PPT states) which are entangled as was first shown in \cite{Pawel} for a qutrit-qutrit system. Hence, one needs more refined methods to check for separability/entanglement.

The most general approach to separability problem is based on the concept of a positive map \cite{Stormer-60,Wor,Stormer,Paulsen} and a directly related concept of an entanglement witness \cite{Terhal}: a linear map $\Phi : \mathcal{B}(\mathcal{H}_1) \longrightarrow  \mathcal{B}(\mathcal{H}_2)$ is positive if $\Phi X \geq 0$ for any $X \geq 0$. In what follows $\mathcal{B}(\mathcal{H})$ denotes a vector space (even a $C^*$-algebra) of bounded operators in $\mathcal{H}$. In this paper we consider only finite dimensional case and hence $\mathcal{B}(\mathcal{H})$ may be viewed as a matrix algebra $M_d(\mathbb{C})$ with $d = {\rm dim}\mathcal{H}$.  A bipartite state $\rho$ is separable iff $(\oper \ot \Phi)\rho \geq 0$ for all positive maps \cite{HHHH}. A Hermitian operator $W$ acting on $\mathcal{H}_1 \ot \mathcal{H}_2$ is an entanglement witness iff $\< \psi_1 \ot \psi_2|W|\psi_1 \ot \psi_2\> \geq 0$ but $W$ is not a positive operator.
The property $\< \psi_1 \ot \psi_2|W|\psi_1 \ot \psi_2\> \geq 0$ is much weaker than the standard positivity of $W$ which is equivalent to $\< \psi |W|\psi\> \geq 0$ for all $\psi \in \mathcal{H}_1 \ot \mathcal{H}_2$. One often calls such operators to be block-positive. Hence, an EW is a block-positive but not positive operator (actually, block-positivity implies that $W$ is Hermitian).
A bipartite state $\rho$ is separable iff ${\rm Tr}(\rho W) \geq 0$ for all entanglement witnesses. Any entangled state may be detected by appropriate positive (but not completely positive) map or by an appropriate entanglement witness \cite{HHHH} (see also \cite{TOPICAL} for the recent review on entanglement witnesses).


Entanglement witness $W$ is called decomposable if $W = A + B^\Gamma$, where  $B^\Gamma = (\oper \ot T)B$ denotes partial transposition, and  $A,B \geq 0$. Such witnesses, however, cannot  detect PPT entangled states, that is, entangled states with Positive Partial Transposition  $\rho^\Gamma \geq 0$. To deal with PPT entangled states one needs non-decomposable witnesses which are much harder to construct. Actually, there is no general construction  of such objects. Moreover, given an EW it is in general very hard to check whether it is decomposable or not.

Another important issue is how much effective is a given witness in detecting entangled states. One  calls $W$ an optimal EW \cite{Lew1,Lew2} if  $W - A$ in no longer block-positive for arbitrary $A\geq 0$, that is $W$ cannot be improved by subtracting a positive operator. A witness $W$ is called nd-optimal \cite{Lew1} if $W - D$ in no longer block-positive for arbitrary decomposable operator $D$. Clearly, nd-optimal witness is necessarily optimal.  It turns out \cite{Lew1,Lew2} that $W$ is nd-optimal if $W$ and $W^\Gamma$ are optimal (authors of \cite{Korea-JMP} call $W$ 1) co-optimal if $W^\Gamma$ is optimal, and 2) bi-optimal if both $W$ and $W^\Gamma$ are optimal). An EW has a {\em spanning property} if there exists a set of product vectors $|\psi_k \> \ot |\phi_k\>$ such that $\< \psi_k \ot \phi_k|W| \psi_k \ot \phi_k \> =0$ spans the Hilbert space $\mathcal{H}_1 \ot \mathcal{H}_2$. Lewenstein et al \cite{Lew1} proved that any witness with spanning property is optimal. It should be stressed, however, that there are optimal witnesses which do not have spanning property.

Finally, since the set of block-positive operators is convex one may consider its extremal elements.  $W$ is extremal if it satisfies the following property:
if $W - A$ is block-positive for some block-positive operator $A$, then $A = aW$ with $a \leq 1$. Among extremal elements there is a dense set of so called exposed elements.    An extremal $W$ is exposed if it satisfies the following property: suppose there exists a separable state $\rho_{\rm sep}$ such that ${\rm Tr}(W \rho_{\rm sep} ) = 0$ and let ${\rm Tr}(W' \rho_{\rm sep} ) = 0$    for some block-positive operator $W'$, then $W' = aW$, with $a >0$. Interestingly, spanning property is sufficient (but not necessary) for optimality and necessary (but not sufficient) of exposedness (cf. Kye review \cite{KYE} devoted to geometric structures related to the set of entanglement witnesses and Hansen {\it et al.} \cite{Trondheim} for the review of extremal entanglement witnesses).

In this paper, we analyse a class of EWs constructed in terms of mutually unbiased bases. It turns out that our construction reproduces many well-known examples of witnesses or equivalently positive maps like the celebrated reduction map and Choi map together with its generalizations. We discuss the issue of optimality and extremality. This problem is very hard, and only partial results are presented.

\section{Positive maps from MUBs}

Let us recall that two orthonormal bases $|\psi_k\>$ and $|\phi_l\>$ in $\Cd$ define mutually unbiased bases (MUBs) iff for any $k$ and $l$ the following condition is satisfied:
\begin{equation}
  |\<\psi_k|\phi_l\>|^2 = \frac 1 d .
\end{equation}
Moreover, it is well-known \cite{Wootters} that the number $N(d)$ of MUBs in $\mathbb{C}^d$ is bounded by $N(d) \leq d+1$ \cite{MAX} (see \cite{MUB-Karol} for the review). If $d=p^r$ with $p$ being a prime number, one has $N(d)=d+1$. In this case, explicit constructions are known \cite{Wootters,MAX}. If $d=d_1d_2$, then $N(d) \geq \min \{N(d_1),N(d_2)\}$ \cite{MUB-1}. Moreover, Grassl \cite{MUB-2} provided a construction of three MUBs in an arbitrary dimension. MUBs already found many important applications in quantum tomography \cite{Wootters,T0,T1}, quantum cryptography \cite{Cr1,Cr2},  the mean king's problem \cite{MK1,MK2}, and entropic uncertainty relations \cite{Entr1,Entr2,Entr3}. Recently, they were used to witness entangled quantum states \cite{Beatrix,W-MUB} and in \cite{Maccone-2} (last result was recently generalized to the multipartite scenario \cite{Maccone}).

Now, we provide the construction of the large class of trace-preserving positive maps in $M_d(\mathbb{C})$: let
$  \{ |\psi^{(\alpha)}_1\>,\ldots,|\psi^{(\alpha)}_{d}\> \}$ with $\alpha=1,\ldots,L$ denote $L$ MUBs (clearly $L \leq N(d) \leq d+1$).  Let us denote the corresponding rank-1 projectors by $P^{(\alpha)}_l = |\psi^{(\alpha)}_l\>\< \psi^{(\alpha)}_l|$. Moreover, let $ \mathcal{O}^{(\alpha)}$ be a set of orthogonal  rotation in $\mathbb{R}^d$ around the axis $\mathbf{n}_*=(1,1,\ldots,1)/\sqrt{d}$, that is, $\mathcal{O}^{(\alpha)} \mathbf{n}_* =  \mathbf{n}_*$.

\begin{Theorem} \label{I} The following map
\begin{equation}\label{Phi}
  \Phi X = \Phi_* X - \frac{1}{d-1}  \sum_{\alpha=1}^{L} \sum_{k,l=1}^d  \mathcal{O}^{(\alpha)}_{kl} {\rm Tr}(\widetilde{X} P^{(\alpha)}_l) P^{(\alpha)}_k ,
\end{equation}
where   $\widetilde{X} = X -  \Phi_* X$ denotes the traceless part of $X$ ($\Phi_* X = \frac 1d \mathbb{I}_d {\rm Tr}X$ denotes the completely depolarizing channel) is positive and trace-preserving.
\end{Theorem}
Proof: denote by $\mathcal{D}(d)$ the space of $d \times d$ density matrices.  Recall that in $\mathcal{D}(d)$ one may inscribe a maximal ball $B_\star$ center at the maximally mixed state $\rho_\star = \frac 1d \mathbb{I}$ \cite{KAROL}: $\rho \in B_\star \subset \mathcal{D}(d)$ iff
\begin{equation}\label{}
  {\rm Tr}\rho^2 \leq \frac{1}{d-1} .
\end{equation}
To prove positivity of $\Phi$ we show that for any rank-1 projector $P=|\psi\>\<\psi|$ one has
\begin{equation}\label{}
  {\rm Tr}(\Phi P)^2 \leq \frac{1}{d-1} ,
\end{equation}
that is, $\Phi$ maps any rank-1 projector into the ball $B_\star$. One finds
\begin{widetext}
\begin{eqnarray*}
  {\rm Tr}(\Phi P)^2 &=& {\rm Tr}\left\{ \frac{1}{d^2} \mathbb{I} - \frac{2}{d-1}\sum_{\alpha=1}^{L} \sum_{k,l=1}^d  \mathcal{O}^{(\alpha)}_{kl} {\rm Tr}(\widetilde{P} P^{(\alpha)}_l) P^{(\alpha)}_k   +  \frac{1}{(d-1)^2} \sum_{\alpha,\beta=1}^{L} \sum_{k,l,m,n=1}^d \mathcal{O}^{(\alpha)}_{kl} {\rm Tr}(\widetilde{P} P^{(\alpha)}_l) P^{(\alpha)}_k  \mathcal{O}^{(\beta)}_{mn} {\rm Tr}(\widetilde{P} P^{(\beta)}_n) P^{(\beta)}_m \right\} \\ &=& \frac 1d -  \frac{2}{d-1}\sum_{\alpha=1}^{L} \sum_{k,l=1}^d  \mathcal{O}^{(\alpha)}_{kl} {\rm Tr}(\widetilde{P} P^{(\alpha)}_l) +
   \frac{1}{(d-1)^2} \sum_{\alpha=1}^{L} \sum_{k,l,m,n=1}^d \mathcal{O}^{(\alpha)}_{kl} \mathcal{O}^{(\alpha)}_{mn} {\rm Tr}(\widetilde{P} P^{(\alpha)}_l)    {\rm Tr}(\widetilde{P} P^{(\alpha)}_n) \delta_{km} \\ &+& \frac 1d \frac{1}{(d-1)^2} \sum_{\alpha\neq \beta=1}^{L}  \sum_{k,l,m,n=1}^d \mathcal{O}^{(\alpha)}_{kl} \mathcal{O}^{(\beta)}_{mn} {\rm Tr}(\widetilde{P} P^{(\alpha)}_l)  {\rm Tr}(\widetilde{P} P^{(\beta)}_n) .
\end{eqnarray*}
\end{widetext}
Now, let us observe that
$$   \sum_{k,l=1}^d  \mathcal{O}^{(\alpha)}_{kl} {\rm Tr}(\widetilde{P} P^{(\alpha)}_l) = 0 , $$
due to $\mathcal{O}^{(\alpha)} \mathbf{n}_* = \pm \mathbf{n}_*$, and hence
\begin{eqnarray*}
  {\rm Tr}(\Phi P)^2 &=& \frac 1d  +  \frac{1}{(d-1)^2} \sum_{\alpha=1}^{L} \sum_{l=1}^d [ {\rm Tr}(\widetilde{P} P^{(\alpha)}_l)]^2 ,
\end{eqnarray*}
where we have used $\sum_k \mathcal{O}^{(\alpha)}_{kl}  \mathcal{O}^{(\alpha)}_{km} = \delta_{lm}  $. Now

$$  [{\rm Tr}(\widetilde{P} P^{(\alpha)}_l)]^2 =  [{\rm Tr}({P} P^{(\alpha)}_l)]^2 + \frac{1}{d^2}- \frac{2}{d}{\rm Tr}({P} P^{(\alpha)}_l) , $$
and using the following inequality \cite{Entr3,Beatrix}

\begin{equation}\label{!}
\sum_{\alpha=1}^{L} \sum_{l=1}^d [{\rm Tr}({P} P^{(\alpha)}_l)]^2 \leq 1 + \frac{L-1}{d} ,
\end{equation}
and finally arrives to

\begin{eqnarray*}
  {\rm Tr}(\Phi P)^2 &\leq & \frac 1d + \frac{1}{(d-1)^2} \left(1 +  \frac{L-1}{d} + \frac{L}{d} - \frac{2L}{d} \right) = \frac{1}{d-1}\\
\end{eqnarray*}
which ends the proof of positivity. The proof of trace-preservation is elementary. \hfill $\Box$

Note, that the formula for $\Phi$ may be rewritten as follows

\begin{widetext}
\begin{equation}\label{Phi}
  \Phi X = \frac{1}{d-1} \left\{ \frac{d+L-1}{d}\, \mathbb{I}\, {\rm Tr}X  - \sum_{\alpha=1}^{L} \sum_{k,l=1}^d  \mathcal{O}^{(\alpha)}_{kl} {\rm Tr}({X} P^{(\alpha)}_l) P^{(\alpha)}_k \right\} ,
\end{equation}
\end{widetext}
and hence it simplifies for $L=d+1$ to the following one
\begin{equation}\label{Phi}
  \Phi X = \frac{1}{d-1} \left\{ 2\, \mathbb{I}\, {\rm Tr}X  - \sum_{\alpha=1}^{d+1} \sum_{k,l=1}^d  \mathcal{O}^{(\alpha)}_{kl} {\rm Tr}({X} P^{(\alpha)}_l) P^{(\alpha)}_k \right\} ,
\end{equation}

Recall that if $L=d+1$, then one may perform complete tomography of $\rho$
\begin{equation}\label{X-P}
  \rho = \frac 1d \mathbb{I}_d + \sum_{\alpha=1}^{d+1} \sum_{k=1}^d a^{(\alpha)}_k P^{(\alpha)}_k ,
\end{equation}
with real parameters
\begin{equation}\label{}
   a^{(\alpha)}_k = {\rm Tr}( \widetilde{\rho}P^{(\alpha)}_k)= {\rm Tr}( {\rho}P^{(\alpha)}_k) - \frac 1d .
\end{equation}
 Hence for each $\alpha=1,\ldots,d+1$ one has
\begin{equation}\label{}
  \sum_{k=1}^d a^{(\alpha)}_k = 0 ,
\end{equation}
which means that the vector $\mathbf{a}^{(\alpha)} =(a^{(\alpha)}_1,\ldots,a^{(\alpha)}_d)$ is orthogonal to the vector $\mathbf{n}_*$.  Having performed complete tomography of $\rho$ one may simplify the proof of Theorem \ref{I}: note that the map $\Phi$ may be rewritten in terms of $a^{(\alpha)}_l$ as follows

\begin{equation}\label{Phi-a}
  \Phi \rho = \Phi_*\rho - \frac{1}{d-1}  \sum_{\alpha=1}^{d+1} \sum_{k=1}^d \sum_{l=1}^d \mathcal{O}^{(\alpha)}_{kl} a^{(\alpha)}_l P^{(\alpha)}_k .
\end{equation}
Now, using $ {\rm Tr} \rho^2 \leq 1$ one finds
\begin{eqnarray}\label{<}
  {\rm Tr} \rho^2 = \frac 1d + \sum_{\alpha=1}^{d+1} \sum_{k=1}^d |a^{(\alpha)}_k|^2 =
   \frac 1d +  \sum_{\alpha=1}^{d+1} |\mathbf{a}^{(\alpha)}|^2 \leq 1 ,
\end{eqnarray}
which implies $\sum_{\alpha=1}^{d+1} |\mathbf{a}^{(\alpha)}|^2  \leq \frac{d-1}{d}$. Finally, using (\ref{<}) and $|\mathcal{O}^{(\alpha)} \mathbf{a}^{(\alpha)}| =  |\mathbf{a}^{(\alpha)}|$, one finds
\begin{equation}\label{}
   {\rm Tr} (\Phi\rho)^2 = \frac 1d + \frac{1}{(d-1)^2}  \sum_{\alpha=1}^{d+1} |\mathbf{a}^{(\alpha)}|^2 \leq \frac{1}{d-1}  ,
\end{equation}
which proves that $\Phi\rho \in \mathbf{B}_* \subset \mathcal{D}(d)$.


Finally, the corresponding entanglement witness

$$ W_\Phi = (d-1)\sum_{i,j=1}^d |i\>\<j| \ot \Phi |i\>\<j| , $$
 reads
\begin{equation}\label{W-Phi}
  W_\Phi= 
    \frac{d+L-1}{d}\, \mathbb{I}_d \ot \mathbb{I}_d -  \sum_{\alpha=1}^{L} \sum_{k,l=1}^d \mathcal{O}^{(\alpha)}_{kl} \overline{P}^{(\alpha)}_l \ot P^{(\alpha)}_k , 
\end{equation}
which simplifies for $L=d+1$ to
\begin{equation}\label{}
  W_\Phi= 
    2\, \mathbb{I}_d \ot \mathbb{I}_d -  \sum_{\alpha=1}^{d+1} \sum_{k,l=1}^d \mathcal{O}^{(\alpha)}_{kl} \overline{P}^{(\alpha)}_l \ot P^{(\alpha)}_k , 
\end{equation}
\begin{Remark} For the maximal set of MUBs, that is, $L=d+1$ the inequality (\ref{!}) is replaced by \cite{Larsen,Ivan}
\begin{equation}\label{!}
 \sum_{\alpha=1}^{d+1} \sum_{l=1}^d [{\rm Tr}({P} P^{(\alpha)}_l)]^2 = 2 ,
\end{equation}
and hence any rank-1 projector $P$ is mapped {\em via} $\Phi$ onto the sphere $S_\star$ being the boundary of $B_\star$.
\end{Remark}

\section{Special classes -- permutations}

The special class of orthogonal $d \times d$ matrices with the additional property $\mathcal{O}\mathbf{n}_* =  \mathbf{n}_*$ is provided by permutations: if $\Pi$ is a permutation matrix then clearly $\Pi \mathbf{n}_* = \mathbf{n}_*$. Taking the simplest case corresponding to $\mathcal{O}^{(\alpha)}=\mathbb{I}_d$ one finds
\begin{equation}\label{}
  \Phi[X] = \Phi_*[X] - \frac{1}{d-1} \sum_{\alpha=1}^{d+1} \sum_{k=1}^d {\rm Tr}(\widetilde{X}P^{(\alpha)}_k)   P^{(\alpha)}_k .
\end{equation}
Now, one easily proves
$$  \sum_{\alpha=1}^{d+1} \sum_{k=1}^d {\rm Tr}(A P^{(\alpha)}_k)   P^{(\alpha)}_k = A + d\Phi_*[A] , $$
and hence
\begin{equation}\label{}
  \Phi[X] = \frac{1}{d-1} \left( \mathbb{I}_d {\rm Tr}X - X\right) ,
\end{equation}
which is the well known reduction map.

Consider now $\mathcal{O}^{(1)} = S$, where $S$ is the permutation defined by $S|i\> = |i+1\>$. Let  $\mathcal{O}^{(2)} =\ldots = \mathcal{O}^{(d+1)} = \mathbb{I}_d$. One finds
\begin{equation}\label{ep}
  \Phi[X] = \frac{1}{d-1} \left( 2 \varepsilon[X] + \sum_{i=2}^{d-1} \varepsilon[S^i X S^{\dagger i}] - X \right) ,
\end{equation}
where
$$\varepsilon[X] = \sum_{i=1}^d P^{(1)}_i X  P^{(1)}_i = \sum_{i=1}^d |i\>\<i|X|i\>\<i| . $$
The map (\ref{ep}) belongs to the family of positive maps
\begin{equation}\label{}
   \tau_{d,k}[X] = \frac{1}{d-1} \left( (d-k) \varepsilon[X] + \sum_{i=1}^{k} \varepsilon[S^i X S^{\dagger i}] - X \right) ,
\end{equation}
developed by Ando \cite{Ando1,Ando2}. Actually, (\ref{ep}) is dual to $\tau_{d,d-2}$.

This construction may be immediately generalized if one considers $d+1$ permutations $\pi^{(\alpha)}$ and defines the  corresponding entanglement witnesses
by
\begin{equation*}\label{}
  W_\Phi=  2\mathbb{I}_d \ot \mathbb{I}_d -  \sum_{\alpha=1}^{d+1} \sum_{k=1}^d \overline{P}^{(\alpha)}_{\pi^{(\alpha)}(k)} \ot P^{(\alpha)}_k  .
\end{equation*}
Actually, these witnesses were analyzed by B. Hiesmayr and A. Rutkowski \cite{Adam}.

\section{A case study:  $d=3$}




For $d=3$ one has four MUBs $ \mathcal{B}_1,\ldots, \mathcal{B}_4$ defined as follows:   $ \mathcal{B}_1 =  \{ \psi^{(1)}_1=|1\>,\psi^{(1)}_2=|2\>,\psi^{(1)}_3=|3\>\}$, where $|1\>,|2\>,|3\>$ defines a computational basis in $\mathbb{C}^3$. The remaining three MUBs are defined as follows
\begin{equation}\label{}
  |\psi^{(\alpha)}_k = U_\alpha |k\> ,
\end{equation}
where the unitary matrices $U_\alpha$ read
\begin{equation*}\label{}
  U_2 = \frac{1}{\sqrt{3}}\left(   \begin{array}{ccc}
   1 &  1 &  1\\
   1 &  \omega^* & \omega  \\
   1 &  \omega & \omega^* \end{array} \right) \ , \ \
   U_3 = \frac{1}{\sqrt{3}}\left(   \begin{array}{ccc}
   1 &  1 &  1\\
   1 &  \omega & \omega^*  \\
   \omega^* &  \omega & 1 \end{array} \right) \ ,
\end{equation*}
and $U_4 = U_3^*$ (with  $\omega=e^{\frac{2 i \pi}{3}}$). One finds for  $\mathcal{B}_2$, $\mathcal{B}_3$ and $\mathcal{B}_4$:

\begin{eqnarray*}
  && \left\{ \frac{|1\> + |2\> +|3\>}{\sqrt{3}},  \frac{|1\> + \omega^* |2\> + \omega|3\>}{\sqrt{3}} ,  \frac{|1\> + \omega |2\> + \omega^* |3\>}{\sqrt{3}} \right\} , \\
   && \left\{ \frac{|1\> + |2\> + \omega^* |3\>}{\sqrt{3}},  \frac{|1\> + \omega |2\> + \omega|3\>}{\sqrt{3}} ,  \frac{|1\> + \omega^* |2\> + |3\>}{\sqrt{3}} \right\} , \\
  && \left\{ \frac{|1\> + |2\> + \omega |3\>}{\sqrt{3}},  \frac{|1\> + \omega^* |2\> + \omega^* |3\>}{\sqrt{3}} ,  \frac{|1\> + \omega |2\> + |3\>}{\sqrt{3}} \right\} .
\end{eqnarray*}
A general proper rotation in $\mathbb{R}^3$ preserving the direction $\mathbf{n}=(n_1,n_2,n_3)$, with $|\mathbf{n}|=1$ is given by the Rodrigues formula
\begin{widetext}
\begin{equation}\label{}
  R(\mathbf{n},\varphi) =  \left(     \begin{array}{ccc}  \cos\varphi + n_1^2(1-\cos\varphi) & n_1 n_2(1-\cos\varphi) - n_3 \sin\varphi & n_1 n_3(1-\cos\varphi) +n_2 \sin\varphi \\  n_1 n_2(1-\cos\varphi) + n_3 \sin\varphi   &  \cos\varphi + n_2^2(1-\cos\varphi) &   n_2 n_3(1-\cos\varphi) - n_1 \sin\varphi    \\
   n_3 n_1(1-\cos\varphi) - n_2 \sin\varphi   &  n_3 n_2(1-\cos\varphi) + n_1 \sin\varphi    &    \cos\varphi + n_3^2(1-\cos\varphi) \end{array} \right) .
\end{equation}
\end{widetext}
Hence taking $\mathbf{n}=\mathbf{n}_*=(1,1,1)/\sqrt{3}$ one finds
\begin{equation}\label{O}
   \mathcal{O}(\varphi) :=  R(\mathbf{n}_*,\varphi) = \left(   \begin{array}{ccc}
   c_1(\varphi) &  c_2(\varphi) & c_3(\varphi) \\
   c_3(\varphi) &  c_1(\varphi) & c_2(\varphi) \\
   c_2(\varphi) &  c_3(\varphi) & c_1(\varphi) \end{array} \right)  \ ,
\end{equation}
where
\begin{eqnarray}   \label{ccc}
  c_1(\varphi) &=& \frac{2}{3}  \cos\varphi + \frac 13 \ , \nonumber \\
  c_2(\varphi) &=& \frac{2}{3}  \cos{\left (\varphi - \frac{2\pi}{3} \right )} + \frac 13\ ,  \\
  c_3(\varphi) &=& \frac{2}{3}  \cos{\left (\varphi + \frac{2\pi}{3} \right )} + \frac 13 \ . \nonumber
\end{eqnarray}
One has $\mathcal{O}(0)=\mathbb{I}_3$. Note that $c_1(\varphi) + c_2(\varphi) + c_3(\varphi) =1$.

In what follows we consider our construction corresponding to $L=4,3,2$ (the case $L=1$ is trivial since one always gets $W \geq 0$ -- actually, in this case there is no sense to use a term MUBs). Interestingly, the corresponding class of maps/witnesses is parameterized by the $L$-dimensional torus.

\subsection{$L=4$}

One finds for the corresponding entanglement witness parameterized by four angles $\{\varphi_1,\varphi_2,\varphi_3,\varphi_4\}$

\begin{equation}\label{ew-abc}
W=\left(
\begin{array}{ccc|ccc|ccc}
 a & \cdot & \cdot & \cdot & p^* & \cdot & \cdot & \cdot & p \\
 \cdot & b & \cdot & \cdot & \cdot & q^* & q & \cdot & \cdot \\
 \cdot & \cdot & c & r^* & \cdot & \cdot & \cdot & r & \cdot \\\hline
 \cdot & \cdot & r & c & \cdot & \cdot & \cdot & r^* & \cdot \\
 p & \cdot & \cdot & \cdot & a & \cdot & \cdot & \cdot & p^* \\
 \cdot & q & \cdot & \cdot & \cdot & b & q^* & \cdot & \cdot \\\hline
 \cdot & q^* & \cdot & \cdot & \cdot & q & b & \cdot & \cdot \\
 \cdot & \cdot & r^* & r & \cdot & \cdot & \cdot & c & \cdot \\
 p^* & \cdot & \cdot & \cdot & p & \cdot & \cdot & \cdot & a \\
\end{array}
\right) ,
\end{equation}
where to make the figure more transparent we replaced all `$0$' by dots. The parameters $\{a,b,c,p,q,r\}$ are defined as follows
\begin{eqnarray*}
a&=&\frac{2}{3} (1-\cos \varphi_1),\\
b&=&\frac{2}{3} \left(\frac{\sqrt{3}}{2} \sin\varphi_1 + \frac{1}{2}\cos\varphi_1 +1\right),\\
c&=&\frac{2}{3} \left(-\frac{\sqrt{3}}{2} \sin\varphi_1 + \frac{1}{2}\cos\varphi_1+1\right),
\end{eqnarray*}
\begin{equation}\label{}
  \left( \begin{array}{c} p \\ q \\ r \end{array} \right) = - \frac 13 \left( \begin{array}{ccc} 1 & 1 & 1 \\ 1 & \omega^* & \omega \\ 1 & \omega & \omega^* \end{array} \right) \left( \begin{array}{c} e^{i\varphi_2} \\ e^{-i \varphi_3}  \\ e^{i \varphi_4}  \end{array} \right) .
\end{equation}

\begin{Example} \label{E1}
Taking $\varphi_2=\varphi_3=\varphi_4=0$ one finds
$$  p=-1,\ \ q=0 ,\ \  r=0 . $$
It reproduces the family of maps analyzed in \cite{KOREA} being generalization of celebrated Choi positive non-decomposable extremal maps \cite{Choi-Lam} (see also \cite{Kossakowski}). It is well known \cite{Korea-PRA,Korea-EXP,Gniewko-OPT} that in this case

\begin{enumerate}

\item $W$ is decomposable iff $b=c$ which means that $\varphi_1 =  0$ or $\varphi_1=\pi $.

\item $W$ is nd-optimal  iff $\varphi_1 \in [-2\pi/3,2\pi/3]$. For $\varphi_1 = \pm 2\pi/3$ one recovers celebrated Choi maps.

\item $W$ has a bi-spanning property iff $\varphi_1 \in (-2\pi/3,2\pi/3)$,

\item $W$ is extremal iff $\varphi_1 \in [-2\pi/3,0) \cup (0,2\pi/3]$,

\item $W$ is exposed iff it is extremal and $\varphi_1 \neq 0$.

\end{enumerate}
For further analysis see also \cite{Filip,DC-JPA}.
\end{Example}

\begin{Example} Taking $\varphi_2=-\varphi_3=\varphi_4=\theta$ one arrives at

\begin{equation}\label{W-theta}
W=   \left(
\begin{array}{ccc|ccc|ccc}
 a & \cdot & \cdot & \cdot & -e^{i \theta} & \cdot & \cdot & \cdot & - e^{-i \theta} \\
 \cdot & b & \cdot & \cdot & \cdot & \cdot & \cdot & \cdot & \cdot \\
 \cdot & \cdot & c & \cdot & \cdot & \cdot & \cdot & \cdot & \cdot \\\hline
 \cdot & \cdot & \cdot & c & \cdot & \cdot & \cdot & \cdot & \cdot \\
 -e^{-i \theta} & \cdot & \cdot & \cdot & a & \cdot & \cdot & \cdot & - e^{i \theta} \\
 \cdot & \cdot & \cdot & \cdot & \cdot & b & \cdot & \cdot & \cdot \\\hline
 \cdot & \cdot & \cdot & \cdot & \cdot & \cdot & b & \cdot & \cdot \\
 \cdot & \cdot & \cdot & \cdot & \cdot & \cdot & \cdot & c & \cdot \\
 -e^{i \theta} & \cdot & \cdot & \cdot & -e^{-i \theta} & \cdot & \cdot & \cdot & a \\
\end{array}
\right) ,
\end{equation}
which was analyzed in \cite{Kye-Theta}.

\end{Example}

\subsection{$L=3$}

Now, one has three orthogonal rotations parameterized by $\{\varphi_1,\varphi_2,\varphi_3\}$. The corresponding operator $W$ has again the structure (\ref{ew-abc}) with the same parameters $a,b,c$, and  the remaining off-diagonal parameters $p',q',r'$ read
\begin{equation}
\left(\begin{array}{c}p' \\q' \\r' \end{array}\right)=-\frac{1}{3}\left(\begin{array}{ccc}1 & 1  \\1 & \omega^* \\ 1 & \omega \end{array}\right)\left(\begin{array}{c}e^{i\varphi_2} \\e^{-i\varphi_3} \end{array}\right) .
\end{equation}

\begin{Example}

Taking $\varphi_2=\varphi_3=0$ one finds

$$  p=- \frac 23 , \ \ q= \frac 13 \omega , \ \  r = q^* . $$
Note, that  in this case taking

$$ a = \frac 43 , \ \ b=c = \frac 13 , $$
one finds $W\geq 0$ which means that the map $\Phi$ is completely positive and hence cannot be used to detect quantum entanglement.
\end{Example}

\subsection{$L=2$}

Now, one has two orthogonal rotations parameterized by $\{\varphi_1,\varphi_2\}$.
Again,  $W$ is given by (\ref{ew-abc}) with the same parameters $a,b,c$, and  the remaining off-diagonal parameters $p'',q'',r''$ read

\begin{equation}
p'' = q'' = r'' =:z =  - \frac 13 e^{i\varphi_2}  .
\end{equation}
One finds

\begin{equation}\label{ew-z}
W=\left(
\begin{array}{ccc|ccc|ccc}
 a & \cdot & \cdot & \cdot & z^* & \cdot & \cdot & \cdot & z \\
 \cdot & b & \cdot & \cdot & \cdot & z^* & z & \cdot & \cdot \\
 \cdot & \cdot & c & z^* & \cdot & \cdot & \cdot & z & \cdot \\\hline
 \cdot & \cdot & z & c & \cdot & \cdot & \cdot & z^* & \cdot \\
 z & \cdot & \cdot & \cdot & a & \cdot & \cdot & \cdot & z^* \\
 \cdot & z & \cdot & \cdot & \cdot & b & z^* & \cdot & \cdot \\\hline
 \cdot & z^* & \cdot & \cdot & \cdot & z & b & \cdot & \cdot \\
 \cdot & \cdot & z^* & z & \cdot & \cdot & \cdot & c & \cdot \\
 z^* & \cdot & \cdot & \cdot & z & \cdot & \cdot & \cdot & a \\
\end{array}
\right) ,
\end{equation}
which is an analog of (\ref{W-theta}). Now, depending upon $\varphi_1$ one may have $W \geq 0$ or $W$ is a proper entanglement witness.

\subsection{PPT entangled state detected by $W$}

Consider the following  $3\ot 3$ state
\begin{equation}\label{ro}
\rho= \frac{1}{15} \left(
\begin{array}{ccc|ccc|ccc}
 1 & \cdot & \cdot & \cdot & 1 & \cdot & \cdot & \cdot & 1 \\
 \cdot & 2 & \cdot & \cdot & \cdot & -1 & -1 & \cdot & \cdot \\
 \cdot & \cdot & 2 & -1 & \cdot & \cdot & \cdot & -1 & \cdot \\\hline
 \cdot & \cdot & -1 & 2 & \cdot & \cdot & \cdot & -1 & \cdot \\
 1 & \cdot & \cdot & \cdot & 1 & \cdot & \cdot & \cdot & 1 \\
 \cdot & -1 & \cdot & \cdot & \cdot & 2 & -1 & \cdot & \cdot \\\hline
 \cdot & -1 & \cdot & \cdot & \cdot & -1 & 2 & \cdot & \cdot \\
 \cdot & \cdot & -1 & -1 & \cdot & \cdot & \cdot & 2& \cdot \\
 1 & \cdot & \cdot & \cdot & 1 & \cdot & \cdot & \cdot & 1 \\
\end{array}
\right) .
\end{equation}
One easily checks that $\rho$ is PPT. Now, taking $\varphi_1=\varphi_2=\pi$ and $\varphi_3=\varphi_4=0$ one finds from (\ref{ew-abc})
\begin{equation}\label{}
W= \frac 13 \left(
\begin{array}{ccc|ccc|ccc}
 4 & \cdot & \cdot & \cdot & -1 & \cdot & \cdot & \cdot & -1 \\
 \cdot & 1 & \cdot & \cdot & \cdot & 2 & 2 & \cdot & \cdot \\
 \cdot & \cdot & 1 & 2 & \cdot & \cdot & \cdot & 2 & \cdot \\\hline
 \cdot & \cdot & 2 & 1 & \cdot & \cdot & \cdot & 2 & \cdot \\
 -1 & \cdot & \cdot & \cdot & 4 & \cdot & \cdot & \cdot & -1 \\
 \cdot & 2 & \cdot & \cdot & \cdot & 1 & 2 & \cdot & \cdot \\\hline
 \cdot & 2 & \cdot & \cdot & \cdot & 2 & 1 & \cdot & \cdot \\
 \cdot & \cdot & 2 & 2 & \cdot & \cdot & \cdot & 1& \cdot \\
 -1 & \cdot & \cdot & \cdot & -1 & \cdot & \cdot & \cdot & 4 \\
\end{array}
\right) ,
\end{equation}
and
\begin{equation}\label{}
  {\rm Tr}(\rho W) = - \frac{2}{15} < 0 ,
\end{equation}
which proves that $\rho$ being PPT is entangled. Interestingly, entanglement of this state is not detected by witnesses from the well known family corresponding to $\varphi_2=\varphi_3=\varphi_4=0$  (cf. Example \ref{E1}). Similarly, one easily checks that $\rho$ is not detected by other three  families of witnesses corresponding to $\varphi_1=\varphi_3=\varphi_4=0$, $\varphi_1=\varphi_2=\varphi_4=0$, and $\varphi_1=\varphi_2=\varphi_3=0$. These are direct generalization of \cite{KOREA} obtained by permuting MUBs. Finally, the realignment test is not conclusive giving the value of realignment $R=1$ (recall that if $R > 1$, then a state is entangled \cite{RR}). To conclude, one cannot detect entanglement of (\ref{ro}) neither using partial transposition/realignment tests nor using previously known entanglement witnesses.

\section{Prime dimensions and Weyl operators and spectra}

Let us recall the construction of Weyl operators \cite{W1,W2,W3}:
$$
U_{kl}=\sum_{k,l=0}^{d-1} \omega^{kl} |m\>\<m+ l| .
$$
They satisfy well known relations
$$ U_{kl}U_{rs}=\omega^{ks}U_{k + r,l + s},\quad U_{kl}^\dag=\omega^{kl}U_{-k,-l} . $$
Authors of \cite{W3} provided the following

\begin{Theorem}\label{Kram}  Let $W$ be a Hermitian operator defined by
\begin{equation}\label{WW}
  W = a \sum_{k,l=0}^{d-1} c_{kl} U_{kl} \ot U_{-k,l} \ ,
\end{equation}
with $ a > 0$ and $c_{00}=d-1$. If the remaining $c_{kl}$ satisfy $|c_{kl}|\leq 1$, then $W$ is a block-positive operator, that is, $\< x \ot y|W|x \ot y\>$ for arbitrary $x,y \in \mathbb{C}^d$.
\end{Theorem}
It is well known that if $d$ is prime then $d+1$ MUBs are directly related to Weyl operators. In this case the set of $d^2-1$ Weyl operators $U_{kl}$ with $(k,l) \neq (0,0)$ splits into $d+1$ sets of mutually commuting operators, that is, $[U_{kl},U_{ij}]=0$ iff $kj = il$ (mod $d$).
These $d+1$ families corresponds to $d+1$ MUBs. Consider as an example $d=3$. One has $U_{00} = \mathbb{I}_3$ and
$$   U_{01} =\left(
\begin{array}{ccc}
 0 & 1 & 0 \\
 0 & 0 & 1 \\
 1 & 0 & 0
\end{array}
\right),\ \   U_{02} =\left(
\begin{array}{ccc}
 0 & 0 & 1 \\
 1 & 0 & 0 \\
 0 & 1 & 0
\end{array}
\right),$$
$$ U_{10}=\left(
\begin{array}{ccc}
 1 & 0 & 0 \\
 0 & \omega & 0 \\
 0 & 0 & \omega^2
\end{array}
\right),\quad U_{11}=\left(
\begin{array}{ccc}
 0 & 1 & 0 \\
 0 & 0 & \omega \\
 \omega^2 & 0 & 0
\end{array}
\right), $$
$$ U_{12}=\left(
\begin{array}{ccc}
 0 & 0 & 1 \\
 \omega & 0 & 0 \\
 0 & \omega^2 & 0
\end{array}
\right), \quad U_{20}=\left(
\begin{array}{ccc}
 1 & 0 & 0 \\
 0 & \omega^2 & 0 \\
 0 & 0 & \omega
\end{array}
\right),$$
$$ U_{21}=\left(
\begin{array}{ccc}
 0 & 1 & 0 \\
 0 & 0 & \omega^2 \\
 \omega & 0 & 0
\end{array}
\right),\quad U_{22}=\left(
\begin{array}{ccc}
 0 & 0 & 1 \\
 \omega^2 & 0 & 0 \\
 0 & \omega & 0
\end{array}
\right) \ ,$$
with $\omega = e^{2\pi i/3}$. One has $d+1=4$ families of commuting operators
$$  \{U_{10},U_{20}\} \ , \  \{U_{11},U_{22}\} \ , \  \{U_{12},U_{21}\} \ , \  \{U_{01},U_{02}\} \ .  $$
One finds that (\ref{WW}) and (\ref{ew-abc}) have the same structure and they are related via: $a=1/3$ together with

\begin{equation}\label{}
  \left( \begin{array}{c} a \\ b \\ c \end{array} \right) = \frac 13  \left( \begin{array}{ccc} 1 & 1 & 1 \\ 1 & \omega^* & \omega \\ 1 & \omega & \omega^* \end{array} \right) \left( \begin{array}{c} c_{00} \\ c_{10}  \\ c_{20}  \end{array} \right) ,
\end{equation}
and

\begin{equation}\label{}
  \left( \begin{array}{c} p \\ q \\ r \end{array} \right) = - \frac 13 \left( \begin{array}{ccc} 1 & 1 & 1 \\ 1 & \omega^* & \omega \\ 1 & \omega & \omega^* \end{array} \right) \left( \begin{array}{c} c_{01} \\ c_{11}  \\ c_{21}  \end{array} \right) .
\end{equation}
Note that hermiticity of $W$ implies that $c^*_{kl} = c_{-k,-l}$ and hence $c_{02} = c_{01}^*$, $c_{11} = c_{22}^*$, $c_{12} = c^*_{21}$. One has therefore
\begin{equation*}\label{}
  c_{10} = e^{i \varphi_1} \ , \ c_{01} = - e^{i\varphi_2} \ , \ c_{22} = - e^{i \varphi_2} \ , \ c_{21} = - e^{i \varphi_4} .
\end{equation*}
that is, $|c_{kl}|=1$ for all pairs $(k,l) \neq (0,0)$.

\section{Conclusions}

We  provided a class of entanglement witnesses/positive maps constructed in terms of Mutually Unbiased Bases (MUBs). Interestingly, this construction reproduces many well known examples like the celebrated reduction map and Choi map together with its generalizations but also gives rise to completely new witnesses/maps. In the 3-dimensional case we obtain a family of witnesses parameterized by 4-dimensional torus. As an example we provided a $3 \ot 3$ entangled state (\ref{ro}) such that one cannot detect its entanglement neither using partial transposition/realignment tests nor using previously known entanglement witnesses.

It is clear that further analysis is needed in order to investigate the issues of optimality and extremality. Such analysis is known only for the special class of Choi-like witnesses (cf. Example \ref{E1}). Also the problem of spanning property deserves further studies. 


Note that if $d=d_1\cdot d_2$ we may consider $\Phi$ as a positive map acting on the matrix algebra of a composite system $M_d(\mathbb{C}) =
M_{d_1}(\mathbb{C}) \ot M_{d_2}(\mathbb{C})$. Interestingly, all density matrices satisfying
$$  {\rm Tr} \rho^2 \leq \frac{1}{d_1d_2-1} , $$
are separable \cite{ZPSM} (actually, they are {\em super-separable}, i.e. separable with respect to arbitrary partition  of $\mathbb{C}^{d_1d_2}$ into a tensor product of $\mathbb{C}^{d_1}$ and $\mathbb{C}^{d_2}$). Hence, they belong to a class of maps analyzed in \cite{barcelona}, that is a class of maps enjoying the following additional property ---  when applied to any state (or a given entanglement class), result in a separable state or, more generally, a state of another certain entanglement class (e.g., Schmidt number $\leq  k$). Another interesting research program may be devoted to further analysis of more general construction of maps in terms of MUBs in the spirit of \cite{barcelona}.

\acknowledgments

DC and GS were supported by the National Science Centre project 2015/19/B/ST1/03095. FW was supported by the Polish Ministry of Science and Higher Education -- "Mobility Plus" Program grant no 1278/MOB/IV/2015/0.

\end{document}